\documentclass[aps,floats,twocolumn,prb,showpacs]{revtex4-1}

\usepackage{amsmath}
\usepackage{graphicx}
\usepackage{xcolor}
\usepackage{epstopdf}
\usepackage{natbib}
\usepackage{xfrac}
\usepackage[normalem]{ulem}
\usepackage{multirow}
\usepackage{bm}
\usepackage{physics}
\usepackage{amssymb}

\usepackage{verbatim}

\newcommand{\intlabel}{\text{int}}

\begin{document}
\title{A quantum Boltzmann equation for strongly correlated electrons}

\author{Antonio~Picano}
\affiliation{Department of Physics, University of Erlangen-N\"urnberg,
	91058 Erlangen, Germany}
\author{Jiajun~Li}
\affiliation{Department of Physics, University of Erlangen-N\"urnberg,
	91058 Erlangen, Germany}
\author{Martin~Eckstein}
\affiliation{Department of Physics, University of Erlangen-N\"urnberg,
	91058 Erlangen, Germany}

\begin{abstract}
Collective orders and photo-induced phase transitions in quantum matter can evolve on timescales which are orders of magnitude slower than the femtosecond processes related to electronic motion in the solid. Quantum Boltzmann equations can potentially resolve this separation of timescales, but are often constructed within a perturbative framework.  Here we derive a quantum Boltzmann equation which only assumes a separation of timescales (taken into account  through the gradient approximation for convolutions in time), but  is based on a non-perturbative scattering integral, and makes no assumption on the spectral function such as the quasiparticle approximation. In particular, a scattering integral corresponding to non-equilibrium dynamical mean-field theory is evaluated in terms of an Anderson impurity model in a non-equilibrium steady state with prescribed distribution functions. This opens the possibility to investigate dynamical processes in correlated solids with quantum impurity solvers designed for the study of non-equilibrium steady states.
\end{abstract}

\date{\today}

\maketitle

\section{Introduction}

One of the biggest challenges in the theoretical description of quantum many-particle systems is to predict their non-equilibrium dynamics at long times after a perturbation. This would be essential for  the understanding of non-equilibrium phenomena in complex solids,\cite{Basov2017, Giannetti2016} including photo-induced  metal-insulator transitions and hidden phases with  spin, orbital, charge, or superconducting order.\cite{Ichikawa2011, Fausti2011a, Beaud2014, Wolf-PRL2014,Stojchevska2014,  Mor2017, Budden2020} 
The evolution of the electronic structure in these situations is often intertwined with the dynamics of the crystal lattice, collective orders, or slow electronic variables such as non-thermal band occupations, which is orders of magnitude slower than intrinsic electronic processes such as the electron tunnelling between atoms.  
Moreover, 
a large timescale separation becomes apparent in the thermalization of pre-thermal states,\cite{Polkovnikov2011, Berges2004,Moeckel2008} where approximate conservation laws provide a dynamical constraint.\cite{Kollar2011, Langen2016}  

A major goal is therefore to devise an approach that can explore the dynamics on the slower timescale, while still taking into account accurately the fast degrees of freedom. Within the Keldysh formalism, non-equilibrium quantum many-particle systems can be described in terms of time and frequency-dependent  spectral functions $A_{\bm k}(\omega,t)$ and  distribution functions $F_{\bm k}(\omega,t)$. (For simplicity,  spin and orbital indices in addition to momentum $\bm k$ are not shown here.)  
A large separation can become evident between the  variation of the functions with time $t$, and the
intrinsic timescales related to the linewidth of relevant spectral features.  
If these timescales are well separated, one can cast the full many-body dynamics into a differential equation known as the quantum Boltzmann equation (QBE).\cite{Kadanoff1962,Kamenev2005} In abstract form, the QBE defines a scattering contribution to the evolution of the distribution functions,
\begin{align}
\label{qbe00}
\big[\partial_t F_{\bm k}(\omega,t)\big]_{scatt} = I[F,A],
\end{align}
where the so-called scattering integral $I$ depends on the 
spectrum
and distribution function at the same time. The full time-dependence is determined by additional contributions from the coherent single-particle propagation,  and a separate equation for the evolution of the 
spectrum
in terms of the distribution function.

While the applicability of the QBE is in principle only controlled by the time-scale separation, the formalism is used  predominantly 
for semiconductors or Fermi liquids with well-defined  quasiparticles,\cite{Haug2008}  
where one can make use of two additional and in practice rather important simplifications: (i) The quasiparticle approximation assumes the spectra $A_{\bm k}(\omega,t)$ to be sharply peaked at energies $\omega=\epsilon_{\bm k}$, and therefore allows to evaluate the QBE on-shell, essentially trading the frequency-dependent distribution function for quasiparticle occupations $n_{\bm k}(t)$. Moreover, (ii), the scattering kernel is often evaluated in a perturbative manner.  In strongly correlated systems, both of these approximations are challenged. For example, doped Mott insulators show strange metallic behaviors without well-defined Fermi liquid quasiparticles in a wide parameter regime \cite{Georges2013AROCMP,Deng2013a}, and similar behavior is  observed  in photo-doped Mott insulators.\cite{Eckstein2013, Sayyad2016,Dasari2020, Petersen2017, Sahota2019} Furthermore, the electronic structure in correlated systems depends strongly on the non-equilibrium distribution, as most clearly demonstrated through the possibility of photo-induced metal insulator transitions. 

For that reason,  the dynamics of correlated systems has been mostly discussed within the formally exact non-equilibrium Green's function (NEGF) techniques. In the NEGF formalism, the dynamics is described in terms of two-time Green's functions $G_{\bm{k}}(t,t')$, which are related to spectra and occupation functions through a Fourier transform with respect to relative time $t-t'$. A two-time self-energy acts as a memory kernel in a non-Markovian propagation of the Green's functions, the so-called Kadanoff-Baym equation.  NEGF techniques can be combined with different diagrammatic approximations,\cite{Golez2016, Babadi2015, Rameau2016, Schluenzen2017} including in particular dynamical mean-field theory (DMFT),\cite{Aoki2014,Georges1996RMP} and  they do not rely on a quasiparticle approximation for the spectrum. On the other hand, they also do not make use of the time-scale separation, and therefore imply a high numerical cost: The effort scales like $\mathcal{O}(t_\text{max}^3)$ with the simulation time $t_\text{max}$, as compared to $\mathcal{O}(t_\text{max})$ for the QBE. For 
weakly interacting systems,  
the perturbatively controlled generalized Kadanoff-Baym Ansatz (GKBA) \cite{Lipavsky1986} has recently been set up to reach $\mathcal{O}(t_\text{max})$ scaling of the computational effort.\cite{Schluenzen2020} For strongly correlated systems, a systematic truncation\cite{Schueler2018} or compact compression\cite{Kaye2020} of the memory kernel in the Kadanoff-Baym equations provide interesting perspectives, but so far the investigation of many fundamental questions has remained out of reach because of the $\mathcal{O}(t_\text{max}^3)$ scaling.

It would therefore be desirable to formulate a QBE which 
incorporates
the simplifications due to the time-scale separation, but does not rely on quasiparticle or perturbative approximations.  
For example, if the equilibrium state of the system is described well by means of DMFT, the steady state fixed point of the QBE should be identical to this DMFT solution.  
A previous work has successfully employed a QBE without the quasiparticle approximation for a Mott insulator,\cite{Wais2018,Wais2020} 
assuming a rigid density of states and a renormalized second-order scattering integral.
Here we show how such a scattering integral can be obtained from an {\em auxiliary non-equilibrium steady state formalism}. This allows to consistently combine the QBE with non-perturbative methods which have been developed to study true non-equilibrium steady states within  DMFT.\cite{Joura2008,Li2015,Titvinidze2018,Matthies2018,Scarlatella2020,Li2020,Panas2019}

The paper is organized as follows: In section \ref{Sec-II}, we present the formulation of a non-perturbative QBE which is consistent with non-equilibrium DMFT. In Sec.~\ref{Sec-III} we compare its solution to a non-equilibrium DMFT simulation for the thermalization in a correlated metal. Section~\ref{Sec-IV} gives a conclusion and outlook. 

\section{Quantum Boltzmann equation}
\label{Sec-II}
\subsection{General setting}

We will derive the QBE for  a generic  model, 
\begin{align}
\label{ham:hubb}
H = \sum_{\bm k ,a,b}h_{\bm k ,ab}(t) c^{\dagger}_{\bm k ,a}c_{\bm k,b}+H_\text{int},
\end{align}
where $c_{\bm k,a}$ ($c_{\bm k,a}^\dagger$) denotes the annihilation (creation) operator for a fermion with spin and orbital indices  $a$ and momentum  $\bm k $, and $H_\text{int}$ is an arbitrary two-particle interaction, and $h_{\bm k ,ab}(t)$ incorporates all 
single-particle terms. We assume that the system of interest is initially prepared in thermal equilibrium at temperature  $T$, and driven out of equilibrium for times $t>0$ by external fields and a coupling to external heat and/or particle reservoirs. The description of this situation within many-body theory is based on contour-ordered  Green's functions,
\begin{align}
\label{GF}
G_{\bm k,ab }(t,t') = -i \langle T_\mathcal{C} c_{{\bm k },a}(t) c_{{\bm k },b}^\dagger(t')\rangle,
\end{align}
with time arguments $t$ and $t'$  on the Keldysh contour $\mathcal{C}$ that runs from 0 to time $t_\text{max}$ (the largest time of interest) on the real time axis, back to 0, and finally to $-i\beta$ along the imaginary time axis. (For an introduction to the Keldysh formalism and the notation, see, e.g., Ref.~\onlinecite{Aoki2014}.) Spin and orbital indices will be no longer shown in the following for simplicity;  all Green's functions, self-energies, dispersion functions  $h_{\bm k}$ are matrices in these indices. From the contour-ordered  function \eqref{GF}, one derives real and imaginary time Green's functions, of which the retarded, lesser, and greater components are most important in the following. The retarded Green's function (with real time arguments)
\begin{align}
\label{GRet}
G^R_{\bm{k}}(t,t')=-i \theta(t-t') \langle[c_{\bm{k}}(t),c^\dagger_{\bm{k}}(t')]_+\rangle,
\end{align}
is related to the spectral function of the system, while the occupied and unoccupied density of states are extracted from the lesser Green's function,
\begin{align}
\label{Gles}
&G^<_{\bm{k}}(t,t')= +i \langle c^\dagger_{\bm{k}}(t') c_{\bm{k}}(t)\rangle.
\\
&G^>_{\bm{k}}(t,t')= -i \langle  c_{\bm{k}}(t)c^\dagger_{\bm{k}}(t')\rangle,
\end{align}
so that  
\begin{align}
\label{Glesretgtr}
G^R_{\bm{k}}(t,t')= \theta(t-t')[G^>_{\bm{k}}(t,t')-G^<_{\bm{k}}(t,t')]. 
\end{align}
In equilibrium, or in any 
time-translationally  
invariant state, all two-time correlation functions depend only on 
the relative time
$t-t'$. By taking the Fourier transform of $G^R$ with respect to this time difference, one obtains the spectral function $A$:
\begin{align}
\label{sepege}
A_{\bm{k}}(\omega)=-\frac{1}{\pi} \Im G^R_{\bm{k}}(\omega+i0),
\end{align}
which is related to the lesser and greater Green's functions through a fluctuation-dissipation theorem
\begin{align}
\label{Fstedy01}
G^<_{\bm{k}}(\omega)&=2 \pi i A_{\bm{k}}(\omega) f_\beta(\omega),
\\
G^>_{\bm{k}}(\omega)&=-2 \pi i A_{\bm{k}}(\omega) [1-f_\beta(\omega)],
\end{align}
where $f_\beta(\omega)$ is the Fermi distribution function, $f_\beta(\omega) =1/(e^{\beta \omega}+1)$. In a non-equilibrium steady-state, one can thus define the distribution function as the ratio
\begin{align}
\label{Fstedy}
	F_{\bm{k}}(\omega)=\frac{G^<_{\bm{k}}(\omega)}{2 \pi i A_{\bm{k}}(\omega)}.
\end{align}
This is an energy distribution function, which is defined even in the absence of well-defined quasi-particles. The QBE provides an equation of motion for its time-dependent generalization, as introduced in the following.

\subsection{The QBE}
\label{sub:BEII}

For every two-time quantity $X(t,t')$ one can introduce the Wigner transform,
\begin{align}
\label{wigner}
X(\omega,t)=\int ds\, e^{i\omega s}\, X(t+s/2,t-s/2),
\end{align}
where $t$ is the average time and 
$s=t-t'$ is the relative time.  
In particular, this can be used to define a time-dependent spectrum and occupation function $F_{\bm k}(\omega,t)$ in analogy to 
Eqs.~\eqref{Glesretgtr}, \eqref{sepege}, and \eqref{Fstedy},
\begin{align}
\label{;xheme}
A_{\bm{k}}(\omega,t)
&=-\frac{1}{\pi} \Im G^R_{\bm{k}}(\omega+i0,t),
\\
&=[G^>_{\bm{k}}(\omega,t)-G^<_{\bm{k}}(\omega,t)]/(-2\pi i),
\\
\label{;xheme01}
F_{\bm k}(\omega,t)
&=
G^<_{\bm{k}}(\omega,t)/[2 \pi i A_{\bm{k}}(\omega,t)],
\end{align}
where $G^{R,<,>}_{\bm{k}}(\omega,t)$ are given by the Wigner transform. 

While Eqs.~\eqref{;xheme} to \eqref{;xheme01} always provide a valid mathematical definition, the functions gain a physical significance in particular in the limit in which there is a well-defined separation of timescales. Let us assume that there are scales $\delta\omega$ and $\delta t$ on which $G(\omega,t)$ varies in frequency and time, such that 
\begin{align}
\label{tsep01}
\Big|\frac{\partial_\omega G_{\bm k}(\omega,t)}{G_{\bm k}(\omega,t)}\Big| < 1/\delta \omega,\,\,\,\,
\Big|\frac{\partial_t G_{\bm k}(\omega,t)}{G_{\bm k}(\omega,t)}\Big|< 1/\delta t,
\end{align}
for lesser, greater, or retarded component. 
The scale $\delta \omega$ measures the relevant internal energy  differences in the system, such as the linewidth or relevant spectral features, 
and $\delta t$  sets 
the scale for the time-evolution, 
with 
$\delta t\to \infty$ in a steady state. The QBE will be derived in the limit where  these timescale are well separated, 
\begin{align}
\label{tsep}
\delta t \gg 1/\delta \omega.
\end{align}
This is also the limit in which the spectral and occupation functions gain their usual meaning in terms of a density of states:
One can always approximate  $G(\omega,t)$ by the average
\begin{align}
\label{averafe}
G(\omega,t)
\!\approx\!\!
\int \frac{dt' d\omega'}{\pi\Omega\tau}  e^{-(\frac{t'}{\tau})^2-(\frac{\omega'}{\Omega})^2}
G(\omega+\omega',t+t')
\end{align}
over a time interval $\tau\ll \delta t$ and  a frequency interval  $\Omega\ll \delta\omega$ on which the function varies weakly. With a sufficiently large time-scale separation \eqref{tsep}, it is possible to choose $\Omega=1/\tau$ without violating the conditions $\tau \ll \delta t$ and $\Omega\ll \delta\omega$. With this, the average \eqref{averafe}, with $G$ replaced by $-iG^<$, is the expression for the time-resolved photoemission spectrum \cite{Freericks2009,Eckstein2008} computed with a Gaussian probe pulse of duration $\tau$, and  therefore has a well-defined interpretation in terms of an occupied density of states. In addition, this  implies that the expression is real and positive, which can be proven by casting Eq.~\eqref{averafe} in the form of a complete square  using a Lehmann representation for the Green's function.  In the same way, $iG^>(\omega,t)$ can be interpreted as the unoccupied density of states (electron addition spectrum), and the spectral function $A(\omega,t)=[G^>(\omega,t)-G^<(\omega,t)]/(-2\pi i)$ has the usual meaning of a single-particle density of states in the many-body system.

The QBE 
provides an equation of motion for the spectral and occupation functions \eqref{;xheme} and \eqref{;xheme01} in the limit of well separated times.\cite{Kamenev2005} Most importantly, the limit \eqref{tsep} allows for the simplification of the convolution $[A\ast B](t,t')=\int d\bar t A(t,\bar t)B(\bar t,t')$ of two real-time functions $A$ and $B$. In mathematical terms, the Wigner transform of the convolution is given by the Moyal product
\begin{align}
\label{moyal}
[A\ast B](\omega,t)
=
e^{\frac{i}{2}[\partial_t^A\partial_\omega^B-\partial_t^B\partial_\omega^A]} A(\omega,t) B(\omega,t).
\end{align}
If  Eqs.~\eqref{tsep01} and \eqref{tsep} hold for $A$ and $B$,  the Moyal product can be simplified by considering only the leading term
\begin{align}
\label{moyalapproy}
[A\ast B](\omega,t)
\approx
A(\omega,t) B(\omega,t),
\end{align}
because $|\partial_t A \: \partial_\omega B| \ll |AB|  $. This is the so-called \textit{gradient approximation}. In a time-evolving state, Eq.~\eqref{Fstedy01} is generalized to the ansatz
\begin{align}
\label{ansatz0001}
G_{\bm k}^<(t,t') = [F_{\bm k} \ast G_{\bm k}^A](t,t') - [G_{\bm k}^R \ast F_{\bm k}](t,t'),
\end{align}
where $F_{\bm k}(t,t')$ depends on two-times, and $G_{\bm k}^A(t,t')= G_{\bm k}^R(t',t)^\dagger$ is the advanced Green's function. By applying the gradient approximation \eqref{moyalapproy} to this ansatz, we  obtain the factorization
\begin{align}
\label{Glestime}
	G^<_{\bm{k}}(\omega,t)=2 \pi i A_{\bm{k}}(\omega,t) F_{\bm{k}}(\omega,t),
\end{align} 
equivalent to  Eq.~\eqref{;xheme01},  using $G^A_{\bm k }(\omega,t)=G^R_{\bm k }(\omega,t)^\dagger$.

In order to derive the QBE for the evolution of the distribution function $F_{\bm k}$, one can consider the equations of motion for the Green's function. For a  non-interacting system with Green's function
$\mathcal{G}_{\bm{k}}(t,t')=-i \langle \mathcal{T}_{\mathcal{C}}c_{\bm{k}}(t)c_{\bm{k}}^\dagger(t') \rangle$
this is written as 
\begin{align}
&\{\mathcal{G}^{-1}_{\bm k} \ast \mathcal{ G}_{{\bm k}}\}(t,t')=\delta_{\mathcal{C}}(t,t'),
\\
&\mathcal{G}^{-1}_{\bm k}(t,t')  = [i\partial_t+\mu-h_{\bm k} (t)]\delta_{\mathcal{C}}(t,t'),
\end{align} 
where $\delta_{\mathcal{C}}(t,t')$ represents the delta-function on the Keldysh contour and $\mu$ is the chemical potential of the system. In the following, it will be convenient to also include the Hartree and Fock self-energy into the dispersion $h_{\bm k}(t)$. To include correlations we take into account  the contour-ordered self-energy $\Sigma(t,t')$ and obtain the interacting Green's function $G$ via the Dyson equation
\begin{align}
\label{Dyson}
\{[(\mathcal{G}_{\bm k})^{-1} - \Sigma_{\bm k}] \ast G_{\bm k}\}(t,t')=\delta_{\mathcal{C}}(t,t')
\end{align} 
on the Keldysh contour. From the Dyson equation for the lesser component, 
[$(\mathcal{G}^R_{ \bm k})^{-1}-\Sigma^R_{\bm k}] \ast G_{\bm k}^< = \Sigma^<_{\bm k}\ast G_{\bm k}^A$,
and the ansatz \eqref{ansatz0001}, we get
\begin{align}
(\mathcal{G}^R_{ \bm k})^{-1} \ast F_{\bm k}
-
&
F_{\bm k} \ast (\mathcal{G}^A_{\bm k})^{-1}
=
\nonumber
\\
&= 
\Sigma^<_{\bm k} +  \Sigma^R_{\bm k} \ast F_{\bm k}  - F_{\bm k} \ast \Sigma^A_{\bm k}.
\end{align}
(Real-time time arguments are shown only where otherwise ambiguous.)
We thus obtain the equation of motion for $F_{\bm k}(t,t')$:
\begin{align}
\label{kwdqbodqwlw}
i(\partial_t + \partial_{t'}) F_{\bm k}(t,t')
=&
h_{\bm k}(t)F_{\bm k}(t,t') - F_{\bm k}(t,t')h_{\bm k}(t')
\nonumber
\\
&+
\Sigma^<_{\bm k} +  \Sigma^R_{\bm k}\ast F_{\bm k}  - F_{\bm k} \ast \Sigma^A_{\bm k}.
\end{align}
Equation \eqref{kwdqbodqwlw} is still exact. To obtain the QBE, we then use the gradient approximation \eqref{moyalapproy} to rewrite Eq.~\eqref{kwdqbodqwlw} as 
\begin{align}
\label{kwdqbodqwlw01}
\partial_t F_{\bm k}
(\omega,t)
&= 
-i[h_{\bm k}(t),F_{\bm k}(\omega,t)]+I_{\bm k}(\omega,t),
\\
I_{\bm k}(\omega,t)
&=
-i\big[
 \Sigma_{\bm k}^R(\omega,t)F_{\bm k}(\omega,t)-F_{\bm k}(\omega,t)\Sigma_{\bm k}^A(\omega,t) 
\nonumber
\\
&\,\,\,\,\,\,\,+\Sigma^<_{\bm k}(\omega,t)\big],
\label{gdgegehe}
\end{align}
where $I_{\bm k}(\omega,t)$ is the \textit{scattering integral.} This equation is completed by the Dyson equation for the retarded Green's function to  leading order in the gradient approximation,
\begin{align}
\label{GR}
G^R_{\bm k}(\omega,t)
&= 
[\omega +i0 +\mu - h_{\bm k}(t) - \Sigma_{\bm k}^R(\omega,t) ]^{-1}.
\end{align}
This set of equations must be combined with a given expression for the self-energy.  
For example, a simple perturbative expression 
would be a second-order diagram in terms of a two-particle density-density interaction $v_{\bm q},$
\begin{align}
\label{sig2}
\Sigma_{\bm k}(t,t')
=
\sum_{\bm k',\bm q} v_{\bm q}^2 G_{\bm k'+\bm q}(t,t')G_{\bm k'}(t',t)G_{\bm k'-\bm q}(t,t').
\end{align}
Such an analytic perturbative expression for $\Sigma$ can then  
be evaluated in the gradient approximation, thus closing the equation. In the following, we discuss a strategy to incorporate a non-perturbative self-energy 
approximation like DMFT into the QBE formalism,
in which an explicit analytical expression for $\Sigma$ is not given.

\subsection{Non-perturbative evaluation of the scattering integral}

In general, the self-energy includes contributions from the interaction, and a possible coupling to a noninteracting environment, which can be used to represent thermal and particle reservoirs \cite{Tsuji2009, Buettiker1985, Aoki2014}. In the following, we write $\Sigma= \Sigma_{\intlabel}+\Gamma$, where $\Sigma_{\intlabel}$ is the interaction contribution, and $\Gamma$ represents the noninteracting reservoirs. Evaluating the interaction self-energy is the main challenge.  We assume that the interaction self-energy $\Sigma_{\intlabel}(t,t')=\hat \Sigma_{\bm k,t,t'}^\text{skel}[G]$ is a functional of the full Green's function $G$, as obtained in particular as the so-called skeleton 
expansion through  
derivatives of the Luttinger-Ward functional\cite{Luttinger1960} for any conserving approximation.\cite{Baym1961} 
Also
DMFT and its extensions can be cast in this language.\cite{Georges1996RMP} A simple perturbative example would be the second-order diagram Eq.~\eqref{sig2}.
Let us now imagine a system which has the same interaction but general non-interacting reservoirs so that the system resides in a non-equilibrium steady state (NESS) with steady state spectrum $\bar A_{\bm k}(\omega)$, and the steady state distribution $\bar F_{\bm k}(\omega)$. Evaluation of the full skeleton functional $\hat \Sigma_{\bm k,t,t'}^\text{skel}[G]$ at the translationally invariant Green's function $\bar G[\bar A,\bar F\big]$ defines a non-equilibrium steady-state functional through the Wigner transform \eqref{wigner}
 \begin{align}
\label{gejen}
\hat\Sigma^\text{ness-skel}_{\bm k,\omega}\big[\bar A,\bar F\big]
=
\int ds\, e^{i\omega s}\,
\hat \Sigma_{\bm k, s/2,-s/2}^\text{skel}[\bar G].
\end{align}
This  
skeleton  functional is universal in the sense that it parametrically depends only on the interaction,\cite{Potthoff2003b} but not on the single-particle part of the Hamiltonian, and hence the functional \eqref{gejen} is independent of the choice of the reservoirs. 
In order to write the equations below in a more compact form, we note that the self-consistent evaluation of the functional \eqref{gejen}, together with the steady state Dyson equation for the retarded function 
\begin{align}
\bar A_{\bm k}(\omega)
&= 
-\frac{1}{\pi}\text{Im}
\frac{1}{\omega^+ +\mu - \bar h_{\bm k} - \bar \Gamma^R_{\bm k}(\omega)-\bar \Sigma^R_{\intlabel,\bm k}(\omega)}
\end{align}
and given $\bar h_{\bm k}$ and $\bar \Gamma^R_{\bm k}(\omega)$,  implicitly defines
a steady-state  functional of the self-energy and the spectral function in terms of the distribution function only, which we will denote by
\begin{align}
\label{gejen02yz}
\hat\Sigma_{\bm k,\omega}^\text{ness}\big[\bar F;\bar h_{\bm k},\bar \Gamma^R_{\bm k}\big],
\,\,\,\,
\hat A_{\bm k,\omega}^\text{ness}\big[\bar F;\bar h_{\bm k},\bar \Gamma^R_{\bm k}\big].
\end{align}

Back to the QBE, at each order of a diagrammatic expression, the two-time self-energy $\Sigma_{\intlabel}(t,t')$ can be written as a sum of convolutions and products of the  full Green's function $G$. In each of these terms, one can consistently use the leading order of the gradient approximation, in combination with the factorization \eqref{Glestime}. This procedure would be the same as evaluating $\hat \Sigma_{\intlabel,\bm k,t,t'}^\text{skel}[\bar G]$ with a time-translationally invariant function $\bar G$ with spectral function $\bar A_{\bm k}(\omega)=A_{\bm k}(\omega,t)$ and distribution function  $\bar F_{\bm k}(\omega)=F_{\bm k}(\omega,t)$. Hence the self-energy in the gradient approximation amounts to evaluating the NESS functional \eqref{gejen}
\begin{align}
\label{ness-skel-t}
\Sigma_{\intlabel,\bm k}(\omega,t)
=
\hat\Sigma^\text{skel-ness}_{\bm k,\omega}\big[A(\cdot,t),F(\cdot,t)\big].
\end{align}
Here the notation $X(\cdot,t)$ of the functional arguments $X=A,F$ indicates that the latter are considered as function of all their arguments except for $t$, which is considered as a fixed parameter.
With Eq.~\eqref{gejen02yz}, the QBE is now formally written as
\begin{align}
\label{kwdqbodqwlw10}
\partial_t F_{\bm k}(\omega,t)
&= 
-i[h_{\bm k}(t),F_{\bm k}(\omega,t)]+I_{\bm k,\omega}[F(\cdot,t)],
\\
\label{kwdqbodqwlw11}
I_{\bm k,\omega}[F(\cdot,t)]
&=
-i\big[
 \Sigma_{\bm k}^R(\omega,t)F_{\bm k}(\omega,t)-F_{\bm k}(\omega,t)\Sigma_{\bm k}^A(\omega,t) 
\nonumber
\\
&\,\,\,\,\,\,\,+\Sigma^<_{\bm k}(\omega,t)\big],
\end{align}
where in the second line $\Sigma=\Sigma_{\intlabel}+\Gamma$, with
\begin{align}
\label{kwdqbodqwlw12}
\Sigma_{\intlabel,\bm k}(\omega,t)
&=
\hat\Sigma_{\bm k,\omega}^\text{ness}\big[ F(\cdot,t); h_{\bm k}(t),\Gamma^R_{\bm k}(\cdot,t) \big].
\end{align}
In addition, the spectral function is given by
\begin{align}
\label{kwdqbodqwlw13}
A_{\bm k}(\omega,t)&=\hat A_{\bm k,\omega}^\text{ness}\big[ F(\cdot,t); h_{\bm k}(t),\Gamma^R_{\bm k}(\cdot,t)\big].
\end{align}
Physically, the last equation \eqref{kwdqbodqwlw13} means that  we allow the electronic distribution function to instantaneously influence the electronic structure of the material. We will therefore refer to Eq.~\eqref{kwdqbodqwlw13} as the \textit{instantaneous response approximation}. 

Equations~\eqref{kwdqbodqwlw10} to \eqref{kwdqbodqwlw13} now provide a closed set of time-dependent equations. This implicit scheme allows a non-perturbative evaluation of the QBE, provided that an efficient numerical description of a NESS is available: To evaluate $\hat\Sigma_{\bm k,\omega}^\text{ness}\big[ F(\cdot,t),...\big]$ and $A_{\bm k}(\omega,t)=\hat A_{\bm k,\omega}^\text{ness}\big[F(\cdot,t),...\big]$ for a given distribution function $\bar F$, we choose an {\em auxiliary steady state system} with reservoir self-energy $\bar \Gamma^R_{\bm k}(\omega)= \Gamma^R_{\bm k}(\omega,t)$, while the bath occupation function, and hence $\bar \Gamma^<_{\bm k}(\omega)$ is treated as a free parameter. The latter is chosen such that the solution $\bar F_{\bm k}(\omega)$ gives the prescribed $F_{\bm k}(\omega,t)$, after which the outcomes $\bar A_{\bm k}(\omega)$ and $\bar\Sigma_{\intlabel,\bm k}(\omega)$ are used to evaluate \eqref{kwdqbodqwlw12} and \eqref{kwdqbodqwlw13}. In particular, within non-equilibrium DMFT, where only local self-energies need to be evaluated in a quantum impurity model, several promising non-perturbative techniques are available that can directly target such non-equilibrium states (see discussion in Sec.~\ref{Sec-IV}). Once Eqs.~\eqref{kwdqbodqwlw12} and \eqref{kwdqbodqwlw13} can be evaluated for a given $F$, the QBE  Eq.~\eqref{kwdqbodqwlw10} can be solved as any differential equation.
(In the implementation below, we use a simple Runge-Kutta algorithm.)

In the following two sections, we will adapt the general formalism to the non-equilibrium DMFT framework. Before that, we conclude this section with a side remark: It is known even in equilibrium that the self-consistent solution of the Dyson equation with a skeleton self energy functional can have multiple unphysical solutions.\cite{Kozik2015} However, a possible multi-valuedness of the functional \eqref{gejen02yz} will not be a problem here. 
The
functions $A_{\bm k}(\omega,t)$, $F_{\bm k}(\omega,t)$, and $\Sigma_{\bm k}(\omega,t)$  evolve continuously as a function of time,
so that even if unphysical  steady-state solutions exist  for a given distribution function, the physical solution is always selected by the requirement of continuity and the initial condition. On the other hand, if the system would evolve as a function of time into a branching point where multiple solutions of Eq.~\eqref{gejen02yz} meet, this would hint at  a rather unconventional dynamical behavior.
For example, in equilibrium it is known that the multi-valuedness of self-consistent perturbation theory is related to vertex singularities \cite{Schaefer2013}, and in the Hubbard model these vertex singularities apparently fall together with the dynamical critical point found in Ref.~\onlinecite{Eckstein2009}.

\subsection{Scattering integral in DMFT}

In the following, we adapt the general QBE framework to non-equilibrium DMFT. Within DMFT, one maps the lattice model \eqref{ham:hubb} onto an effective single-site impurity model. The impurity site has the same interaction as a site in the lattice, and its coupling to the environment is described by the so-called \textit{hybridization} function $\Delta(t,t')$, which is self-consistently determined such that the local 
($\bm k$-averaged) lattice Green's function
\begin{align}
G_{\text{loc}}(t,t')=\sum_{\bm k}G_{\bm k}(t,t')
\end{align}
coincides with the impurity Green's function. The key approximation of DMFT is that the lattice self-energy is local in space (independent of $\bm k$), 
and one requires the local lattice self-energy to be identical to the impurity self energy. In detail, the impurity model is defined by an action 
\begin{align}
\mathcal{S}
=
-i\int_\mathcal{C} dt\, H_{loc}(t)-i
\int_\mathcal{C} dt dt' 
\sum_\sigma
c_\sigma^\dagger(t) \Delta(t,t')c_\sigma(t'),
\end{align}
in terms of the self-consistent hybridization function. The non-interacting Green's function $\mathcal{G}$ is determined by the Dyson equation
\begin{align}
\label{Dimp1}
\mathcal{G}^{-1}(t,t')=[i\partial_t +\mu -h(t)]\delta_{\mathcal{C}}(t,t') -\Delta(t,t'),
\end{align}
where $h(t)$ is the single particle Hamiltonian in the impurity model. 
The interacting impurity Green's function is given by
\begin{align}
\label{Dimp2}
G_{\text{imp}}^{-1}=\mathcal{G}^{-1}-\Sigma_{\text{imp}},
\end{align}
and the self-consistency requires  
\begin{align}
G_{\text{imp}}=G_{\text{loc}}, 
\,\,\,\,
\Sigma_{\text{imp}}=\Sigma.
\end{align}

The self-consistent impurity model provides an implicit way to evaluate a non-perturbative expression $\hat\Sigma_{\intlabel}[G_{\text{loc}}]$ for a local self-energy in terms of a local Green's functions. Along the line of the previous section, we can therefore use an impurity model in a NESS to construct the steady state functional \eqref{kwdqbodqwlw12} for the local self-energy. An impurity model in the steady state simply implies that the hybridization function itself is translationally invariant in time, and specified through its retarded and lesser components, $\Delta^R(\omega)$ and $\Delta^<(\omega)$. 

The evaluation of the functionals \eqref{kwdqbodqwlw12} and \eqref{kwdqbodqwlw13} within DMFT,  for a given distribution function $\bar F_{\bm k}(\omega)$, depends on the type of impurity solver. 
Below we exemplify this 
for an impurity solver which determines the self energy from an expansion in terms of the noninteracting impurity Green's function $\bar{ \mathcal{G}}$, (such as weak-coupling Keldysh quantum Monte Carlo or iterated perturbation theory):
\begin{itemize}
	\item[1)]
	Start with some guess for $\bar \Sigma_{\intlabel}^R(\omega)$ and $\bar \Sigma_{\intlabel}^<(\omega)$, and calculate the $\bm k$-dependent lattice Green's functions [Eq.~\eqref{GR} with $\bm k$-independent self-energy]
	\begin{align}
	& \bar G_{\bm k}^R(\omega)
	= 
	[\omega +\mu- \bar h_{\bm k} -  \bar\Gamma_{\bm k}^R(\omega)- \bar \Sigma_{\intlabel}^R(\omega)]^{-1}.
	\end{align}
	and the spectrum $\bar A_{\bm k}(\omega)=-\frac{1}{\pi}\text{Im}G_{\bm k}^R(\omega+i0)$.
	\item[2)]
	Determine the lesser Green's function from the given distribution function,
	\begin{align}
	&\bar G_{\bm k}^{<}(\omega) = 2\pi i \bar F_{\bm k}(\omega) \bar A_{\bm k}(\omega).
	\end{align}
	\item[3)]
	Calculate the local lattice Green's functions.
	\begin{align}
	\bar G_\text{loc}^{R,<}(\omega)
	&=
	\sum_{\bm k}
	\bar G_{\bm k}^{R,<}(\omega,t).
	\end{align}
	\item[4)]
	Express the noninteracting Green's function $\mathcal{G}$ of the impurity model in terms of $\Sigma_{\text{imp}}$ of  $G_{\text{imp}}$ using the Dyson equation for the impurity model [Eqs.~\eqref{Dimp1} and \eqref{Dimp2}] in the steady state. For example, this can be written as
	\begin{align}
	\label{qlqlql01}
	&\mathcal{G}^R(\omega)
	=
	[G_{\text{imp}}^R(\omega)^{-1} +  \Sigma_{\text{imp}}^R(\omega)]^{-1},
	\\
	&\Delta^<(\omega)
	=
	G^{R}_{\text{imp}}(\omega)^{-1}
	G^<_{\text{imp}}(\omega)
	G^A_{\text{imp}}(\omega)^{-1}
	-
	 \Sigma^<_{\text{imp}}(\omega)	
	 \nonumber
	\\
		\label{qlqlql03}
	&\mathcal{G}^<(\omega) = \mathcal{G}^{R}(\omega) \Delta^<  \mathcal{G}^A(\omega),
	\end{align} 	
	Solve these equations for $\mathcal{G}(\omega)$ using the DMFT self-consistency for the lattice and impurity quantities,  $  \Sigma_{\text{imp}}(\omega)=\bar \Sigma_{\intlabel}(\omega)$ and 
	$ G_{\text{imp}}(\omega)=\bar G_\text{loc}(\omega)$.  
	\item[5)]
	Calculate a new $\Sigma_{\text{imp}}$ by using an expansion in $\mathcal{G}^{R}(\omega)$. 
	\item[6)]
	Set $\bar \Sigma^{R,<}_{\intlabel}(\omega)=\Sigma^{R,<}_{\text{imp}}(\omega)$, and iterate Step 2) to 5) until convergence.
\end{itemize}
This iteration is 
basically a steady-state non-equilibrium DMFT simulation where the distribution function of the system is prescribed and the distribution of the reservoirs is determined, in contrast to conventional steady-state DMFT where the distribution function of the system of the system is determined by reservoirs with a given distribution function.

\section{Comparison to the full DMFT simulation}
\label{Sec-III}

\subsection{Model}
\label{model-A}

As a first test case for  
the methodology, we study the particle-hole symmetric single-band Hubbard model
\begin{align}
\label{eqa1}
\hat H=  -t_h \sum_{\langle i,j\rangle,\sigma}c^{\dagger}_{i \sigma}c_{j\sigma} + U\sum_{j} \big(\hat{n}_{j\uparrow}-\tfrac12\big)\big(\hat{n}_{j\downarrow}-\tfrac12\big).
\end{align}
Here $c_{j,\sigma}$ denotes the annihilation  operator for a Fermion with spin $\sigma \in \{\uparrow,\downarrow\}$  at lattice site $j$, $\hat{n}_{j\sigma}=c^{\dagger}_{j\sigma}c_{j\sigma}$ is the particle number operator, 
$t_h$ the hopping matrix element between nearest neighbour sites, and $U$ the on-site interaction strength. The actual simulations assume a semi-elliptic local density of states $D(\epsilon)=\sqrt{4-\epsilon^2}/(2\pi)$ for the noninteracting model with bandwidth $4$, corresponding to a Bethe lattice with hopping $t_h=1$. The latter sets the unit of energy, and its inverse defines the unit of time ($\hbar=1$). 

The system is studied in the metallic regime, where $U$ is smaller than the bandwidth. Initially, the system is in equilibrium with a inverse temperature $\beta$. Within a short time interval, we then create  a non-thermal population of electrons and holes similar to a photo-excited population (the precise protocol is given below). This non-thermal population will then relax under the influence of the electron-electron interaction and the coupling to a phonon bath, and we compare a  simulation of this relaxation dynamics within the full non-equilibrium  DMFT simulation and the QBE.

For the excitation, we shortly couple a fermionic reservoir with density of states
\begin{align}
\label{aboth}
A_{\text{bath}}(\omega) =A(\omega-2.5)+A(\omega+2.5)
\end{align}
consisting of two smooth bands with bandwidth $W_{\text{bath}}=6$ around the energies $\omega=\pm2.5$; we choose $A(\omega)=\frac{1}{\pi}\cos[2](\pi\omega/W_{\text{bath}})$ in the interval $[-W_{\text{bath}}/2,W_{\text{bath}}/2]$, see dashed line at the bottom of  Fig~\ref{fig01}c for $A_{bath}(\omega)$. 
Choosing a population inversion in this reservoir will lead to a rapid transfer of electrons from the system into the negative energy part of the reservoir, and of electrons from the positive energy part of the bath to the system, thus generating an electron transfer similar to a photo-excitation process. 
The bath adds a local contribution $\Gamma(t,t')$ to the self-energy (as obtained by integrating out the bath),
\begin{align}
\label{sphonto}
\Gamma(t,t')=V(t)G_{\text{bath}}(t,t')V(t')^*,
\end{align}
where $V(t)$ is the  time-profile of the coupling, and $G_{\text{bath}}(t,t')$ is the bath Green's function,
\begin{align}
G_{\text{bath}}^R(t,t')
&=
-i
\theta(t-t')
\int d\omega \,e^{-i\omega(t-t')}A_{\text{bath}}(\omega),
\\
G_{\text{bath}}^<(t,t')
&=
i
\int d\omega\,
e^{-i\omega(t-t')} f_{\text{bath}}(\omega)A_{\text{bath}}(\omega).
\end{align}
The bath occupation $f_{\text{bath}}(\omega)=f_{-\beta}(\omega)$ is taken to be, during the whole time-evolution of the system, a negative temperature Fermi-Dirac distribution (population inversion) , and the switching profile $V(t)=0.75\sin[2](\pi/5(t-t_0))$ is centred around an early time $t_0=27.5$ with a duration of just five inverse hoppings. 
In general, the QBE is expected to describe the evolution of the system only on timescales much longer than the inverse hopping, so that these details of the excitation protocol are not important for the present study. 

The coupling to the bosonic bath is included via a local electron-phonon self-energy $\Sigma_{\text{ph}}$. In order for the bosons to act as heath bath, we need to neglect the back-action of the electrons on the phonons, and we take $\Sigma_{\text{ph}}$ to be the simple first-order diagram of a local electron-phonon interaction,
\begin{align}
\label{sphonon}
\Sigma_{\text{ph}}(t,t')&=g^2G(t,t') D_{\text{ph}}(t,t'), 
\end{align} 
where $G$ is the fully interacting local electron Green's function of the system, $g$ measures the electron-phonon coupling strength, and $D_{\text{ph}}$ is the propagator for free bosons with an Ohmic density of states $\frac{\omega}{4\omega_{\text{ph}}^2} \exp(-\omega/\omega_{\text{ph}})$  with exponential cutoff $\omega_{\text{ph}}=0.2$. The occupation function of bosons is kept in equilibrium with inverse temperature $\beta$. The temperature of the heat bath is the same as the initial one of the system in equilibrium, such that the system will eventually thermalize back to its initial temperature long after the excitation.

\subsection{Full DMFT solution}

For the semi-elliptic density of states, the  DMFT self-consistency can be formulated in closed form, and the hybridization of the impurity model is simply given by \cite{Georges1996RMP,Aoki2014}
\begin{align}
\label{bethe}
\Delta(t,t')=G(t,t')+\Gamma(t,t')
\end{align}
in terms of the local Green's function $G$. With the non-interacting Green's function of the impurity model
[Eq.~\eqref{Dimp1}], the Dyson equation for the impurity model reads
\begin{align}
\label{dysonimp}
&G^{-1}(t,t')
=
\mathcal{G}^{-1}(t,t')
- \Sigma_{\intlabel}(t,t').
\end{align}
Here 
\begin{align}
\Sigma_{\intlabel}(t,t')
=
\Sigma_{U}(t,t')+\Sigma_{\text{ph}}(t,t')
\end{align}
is the interaction self-energy due to the electron phonon interaction and the Hubbard interaction. The latter is determined using the iterated perturbation theory (IPT) impurity solver, i.e., a second-order expansion in terms of $\mathcal{G}$,
\begin{align}
\Sigma_U(t,t')
=
U^2
\mathcal{G}(t,t')
\mathcal{G}(t,t')
\mathcal{G}(t',t).
\label{IPT}
\end{align}
In addition, the local energy $h(t)$ in Eq.~\eqref{Dimp1} is the Hartree self-energy, $h(t)=U n_{\sigma}(t)$ with the density $n_{\sigma}(t)$  per spin. In the present case we 
study a half-filled system, so that $\mu=U/2$ and $\mu+h(t)=0$.  

The self-consistent solution of the system of Eq. \eqref{bethe} to \eqref{IPT} together with the excitation and phonon self energies Eq.~\eqref{sphonon} and Eq.~\eqref{sphonto} determines the time evolution of the physical system. The equations are solved on the Keldysh contour using the NESSi simulation package.\cite{Schueler2020} For the comparison with the QBE, the local spectral function and distribution function are then extracted from the  Wigner transform of the local Green's function
\begin{align}
A(\omega,t)
&=
-\frac{1}{\pi} \text{Im} G^R(\omega+i0,t),
\\
F(\omega,t)
&=
\frac{G^<(\omega,t)}{2\pi i A(\omega,t)}.
\end{align}
Furthermore, we compute the total energy as:
\begin{align}
\label{energyfull}
E_\text{DMFT}
= -2i(\Delta*G)^<(t,t)- i (\Sigma_{\intlabel}*G)^<(t,t)
\end{align}
The first and second term represent the kinetic and interaction energy, respectively, with a factor two in the kinetic energy for the summation over spin components.

\subsection{QBE formulation}
\label{local}

For the present model, for which a closed set of equations is given in terms of local (momentum-averaged) quantities, the QBE can be derived directly for the local quantities. Instead of deriving Eq.~\eqref{kwdqbodqwlw01} and \eqref{gdgegehe} from the lattice Dyson equation \eqref{Dyson}, one can perform an analogous argument directly for the Dyson equation  of the DMFT impurity model [Eq.~\eqref{dysonimp}]. This leads to a local QBE
\begin{align}
\label{QBE-local}
\partial_t F
(\omega,t)
=&\, 
I[F(\cdot)],
\\
\label{ISigma_local}
I[F(\cdot)]
=&
-i\big(\Sigma^<(\omega,t) + [\Sigma^R(\omega,t)+\Delta^R(\omega,t)]F(\omega,t)+
\nonumber
\\
&-F(\omega,t)[\Sigma^A(\omega,t)+\Delta^A(\omega,t)] \big),
\end{align}
where again $\Sigma=\Gamma+\Sigma_{\intlabel}$, and 
\begin{align}
\label{ness-local-dmft}
\Sigma_{\intlabel}(\omega,t)=&
\Sigma_\omega^{\text{ness}}[F(\cdot,t)],\,\,
A(\omega,t)=
A_\omega^{\text{ness}}[F(\cdot,t)],
\end{align}
$\Sigma(\omega,t)$ and the spectrum $A(\omega,t)$ are understood in terms of an auxiliary steady state impurity model with given prescribed distribution function $\bar F(\omega)=F(\omega,t)$. The evaluation of these functionals is again done iteratively:
\begin{enumerate}
	\item [1)]
	Start from a guess for $\bar \Sigma_{\intlabel}(\omega)$. Solve the steady state variant of Eq.~\eqref{dysonimp} for $\bar G^R(\omega)$, 
	\begin{align}
	& \bar G^R(\omega)
	= 
	[\omega +\mu- \bar h - \Delta^R(\omega)- \bar \Sigma_{\intlabel}^R(\omega)]^{-1}.
	\end{align}
        and determine $\bar A(\omega)=-\frac{1}{\pi} \bar G^R(\omega+i0)$.
	\item[2)]
	Determine the lesser Green's function from the given distribution function, 
	$\bar G^<(\omega) = 2\pi i \bar F(\omega)\bar A(\omega)$.
	\item[3)]
	Use the self-consistency Eq.~\eqref{bethe} to fix the hybridization function of the effective steady state impurity model, $\Delta(\omega)=\bar G(\omega)+\Gamma(\omega)$.
	\item [4)]
	Solve the impurity model. With IPT  as an impurity solver, we first determine $\mathcal{G}(\omega)$ from $\Delta(\omega)$,
	\begin{align}
	& \mathcal{G}^R(\omega)=  [\omega+\mu- h(t) - \Delta^R(\omega) ]^{-1},\\
	&\mathcal{G}^<(\omega) = \mathcal{G}^{R}(\omega)\Delta^< (\omega)  \mathcal{G}^A(\omega),
	\end{align}
	transform  
	to real time, evaluate Eq.~\eqref{IPT}, and transform  
	back to frequency space to obtain $\Sigma_{U}^{R,<}(\omega)$. Similarly, $\Sigma_{\text{ph}}^{R,<}(\omega)$ is evaluated.
	\item[5)]
        Set $\bar \Sigma_{\intlabel}(\omega)=\Sigma_{U}(\omega)+\Sigma_{\text{ph}}(\omega)$, and iterate step 2) to 5) until convergence.
\end{enumerate}
The iteration serves as a way to evaluate $\Sigma^{\text{ness}}[F(\cdot,t)]$. The differential equation \eqref{QBE-local} is 
then solved  using a Runge-Kutta algorithm.  
In addition to the spectral and distribution functions, 
we then compute the  total energy 
\begin{align}
E_\text{QBE}
=-2i [\Delta(\omega)G(\omega)]^<- i [\Sigma_{\intlabel}(\omega)G(\omega)]^<
\label{energy_QBE}
\end{align}
 in order to compare with the full solution \eqref{energyfull}.

\subsection{Results and Discussion}

\begin{figure*}[tbp]
	\centerline{\includegraphics[width=1.0\textwidth]{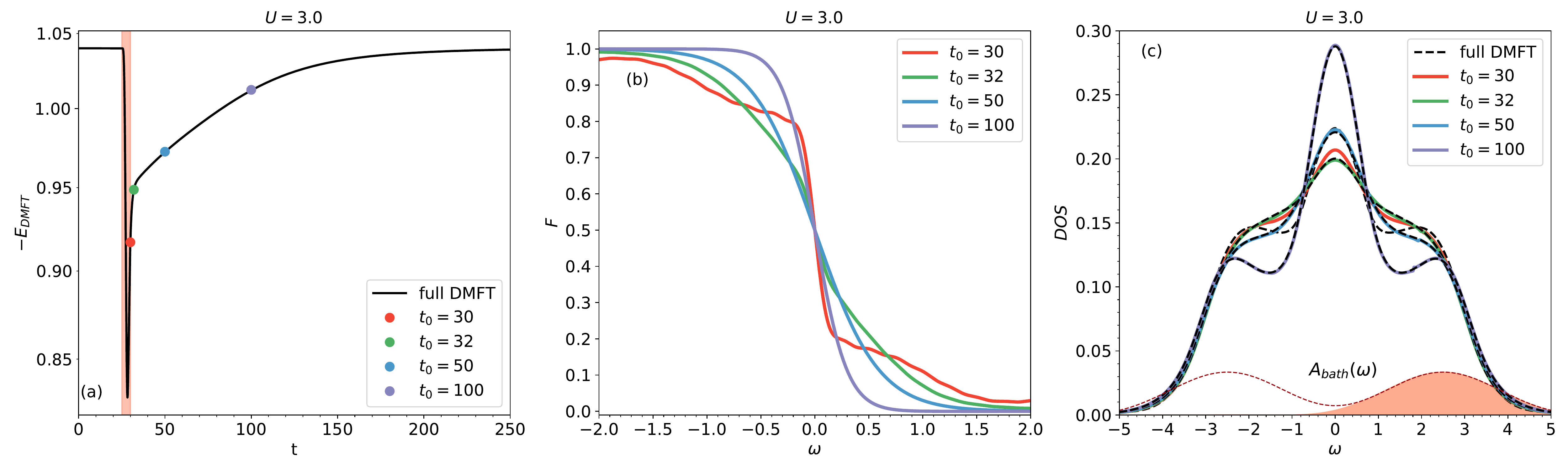}}
	\caption{
a) Energy 
$-E_\text{DMFT}$  
obtained from the full DMFT solution (interaction $U=3$, initial inverse temperature $\beta=20$, electron phonon coupling $g^2=0.5$). 
Coloured dots indicate the energies obtained from the auxiliary steady state $A_\omega^\text{ness}[F] $ [Eq.~\eqref{ness-local-dmft}], for different initial times $t_0$ at which the distribution functions $F(\omega,t_0)$ is taken from the DMFT solution and copied in the auxiliary steady state problem. 
b)  Distribution functions $F(\omega,t_0)$  obtained from the full DMFT solution, at times $t_0$ corresponding to the dots in a), and copied in the auxiliary steady state problem.
c) Dashed lines show the spectrum $A(\omega,t_0)$ at various initial times, obtained from the full DMFT solution. Solid lines 
show the spectra obtained from the auxiliary steady state $A_\omega^\text{ness}[F]$ [Eq.~\eqref{ness-local-dmft}], evaluated with the distribution functions $F(\omega,t_0)$ in b) taken from the DMFT solution. 
The dotted line at the bottom of c) shows the (rescaled) spectral function $A_\text{bath}(\omega)$ [Eq.~\eqref{aboth}] for the excitation bath (the shaded orange area shows the occupied density of states for the bath), and the shaded area in a) the time window over which this bath is coupled to the system.
}
\label{fig01}
\end{figure*}

In this subsection,  we compare the QBE description with the full solution of the KB equations for the setting introduced in Sec.~\ref{model-A}. 
Figure~\ref{fig01}a) 
 shows the evolution of the energy in the full DMFT solution, which increases during the short excitation window, and subsequently relaxes back to the initial state due to electron thermalization and 
 the 
 electron-phonon interaction. 
Figure~\ref{fig01}b) and c)  
  then show the spectra and distribution functions at some points in time. In the initial and final state the spectrum 
  has 
  a central peak, representing a band of renormalized quasiparticles, which coexists with two Hubbard bands around $\omega= \pm U/2$. In equilibrium, with increasing $T$, the quasiparticle peak would be replaced by a dip in the spectral function, indicating that the high-temperature state is a bad-metal without  coherent quasiparticles. After the excitation, the distribution function is highly non-thermal, and the quasiparticle band is strongly suppressed. With time, $F(\omega,t)$ approaches back the shape of an approximate Fermi distribution (electron thermalization), and simultaneously the 
  effective temperature of this distribution relaxes back to  the initial $1/\beta$. Together with 
  this
  evolution of the distribution function, the quasiparticle peak in the spectrum is reformed.

Before computing 
the time evolution generated by the QBE, we can independently 
evaluate
the quality of the  auxiliary steady-state representation of the spectra at each given time, i.e., the accuracy of the functional $A_\omega^\text{ness}[F]$, Eq.~\eqref{ness-local-dmft}: We take  the distribution function $F(\omega,t_0)$ from the full solution at a given time $t_0$, evaluate $A_\omega^\text{ness}[\bar F]$ with $\bar F(\omega)=F(\omega,t_0)$ as described below Eq.~\eqref{ness-local-dmft} to compute a steady state spectrum $\bar A(\omega)$, and compare the result with the full solution  $A(\omega,t_0)$. In Fig.~\ref{fig01}c, dashed lines correspond to the DMFT solution $A(\omega,t_0)$, while solid lines show the corresponding $\bar A(\omega)$. The comparison is perfect, even for relatively early times. Only for times
immediately after the ultrafast excitation ($t=30$),  
where the gradient approximation is not supposed to work, can one observe a 
failure
of the auxiliary steady state representation. We can therefore affirm that the density of states can be very accurately obtained as a steady state functional of the distribution function, even in the correlated metallic regime. For smaller values of $U$, the agreement is as good (not shown here).
Furthermore, not only the density of states can be very accurately obtained as a steady state functional of the distribution function, but the whole Green's function and self-energy: The energy values represented by coloured dots in 
Fig.~\ref{fig01}a), calculated with Eq. \eqref{energy_QBE}, 
exactly match the ones of the full DMFT code at the same time, calculated with 
Eq.~\eqref{energyfull}.  

\begin{figure*}[tbp]
\centerline{\includegraphics[width=1.0\textwidth]{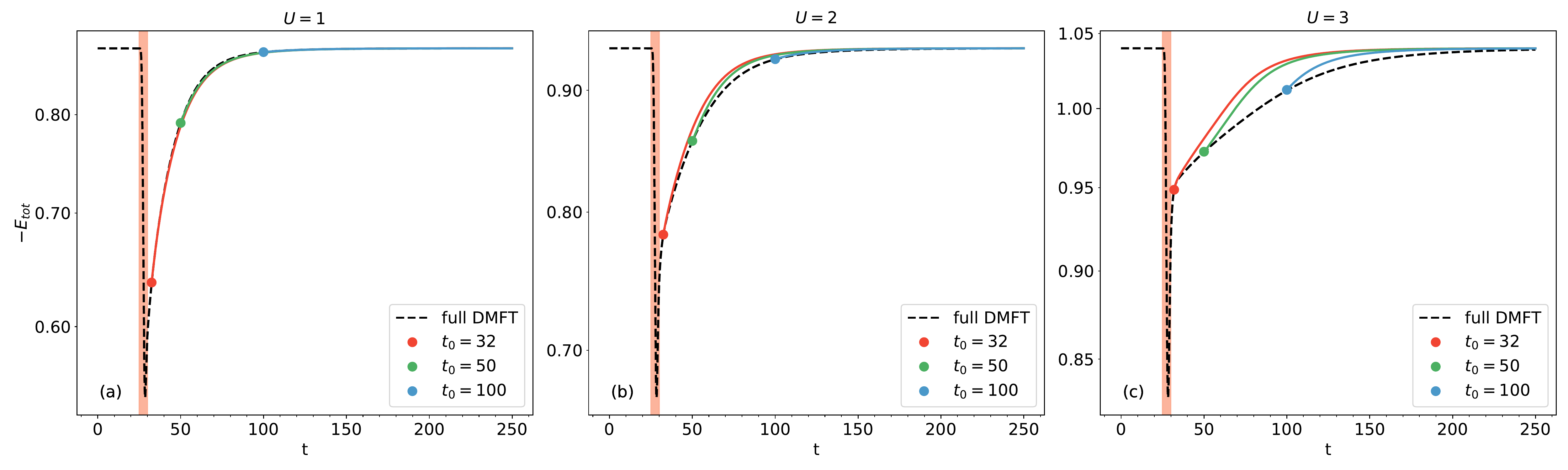}}
\caption{Time-evolution of the total energy for 
$U=1$ (a), $U=2$ (b), and $U=3$ (c) (initial inverse temperature $\beta=20$, electron phonon coupling $g^2=0.5$). The black dashed lines show the energy $-E_\text{DMFT}$ obtained from the full DMFT evolution, 
solid
lines show the energy $-E_\text{QBE}$ obtained from the QBE. The QBE is started at different times $t_0$ (indicated by the dots at the beginning of the dashed lines), taking the distribution function $F_\text{DMFT}(\omega,t_0)$ as an initial state for a solution of the QBE at times $t>t_0$.}
	\label{fig02}
\end{figure*}

In passing, we note that a non-equilibrium spectral function $A(\omega,t)$ defined by the Wigner transform \eqref{wigner} is real (hermitian) by construction, but not necessarily positive, while a steady-state fermionic spectral function is always positive. Moreover, for numerical reasons, for short times the integral in the Wigner transform  \eqref{wigner} is truncated, possibly leading to small artefacts. In practice, the relation $F(\omega,t)=\bar F(\omega)$ will therefore not be enforced exactly, but as a best fit. It should be noted, 
however,
that the positivity of $A(\omega,t)$ and $F(\omega,t)$ is indeed satisfied wherever the gradient approximation is accurate, as discussed in connection with Eq.~\eqref{averafe}. 
In particular, as one can see from Fig.~\ref{fig01}b), the distribution functions are already positive in the relevant time interval for the present case.

Next, we compare the relaxation dynamics of the system in the two descriptions. For this, we simply take the distribution function $F(\omega,t_0)$ at a given time $t_0$ from the full DMFT solution as an initial state for a solution of the QBE for $t>t_0$. The time-evolution of the energy is shown in Fig.~\ref{fig02} for three different values of $U$, and different starting times $t_0$ of the QBE simulation. 
For small values of $U$ ($U=1$ and $U=2$ in 
Fig.~\ref{fig02}a) 
and b), respectively), the energy relaxation rate obtained from the QBE is almost identical to the one from full DMFT.
For $U=3$ 
(Fig.~\ref{fig02}c),  
one can observe a difference in the magnitude of the time-constants related to the relaxation of the total energy in the two approaches.
In particular, the 
QBE 
presents an artificially faster relaxation with respect to the full DMFT 
solution.
This indicates that the gradient approximation is less justifies for $U=3$, which could be related to the existence of a more narrow quasiparticle band. As the starting point $t_0$ of the Boltzmann code shifts forward in time, the difference between the time evolution of the energies becomes less pronounced.
If one decreases the coupling $g^2$ with the phonon bath (not shown), the relaxation dynamics of the system is slowed down, the gradient approximation is more justified, and the difference in the energy relaxation rate in the
two
approaches is less pronounced.

\begin{figure}[tbp]
	\centerline{\includegraphics[width=0.5\textwidth]{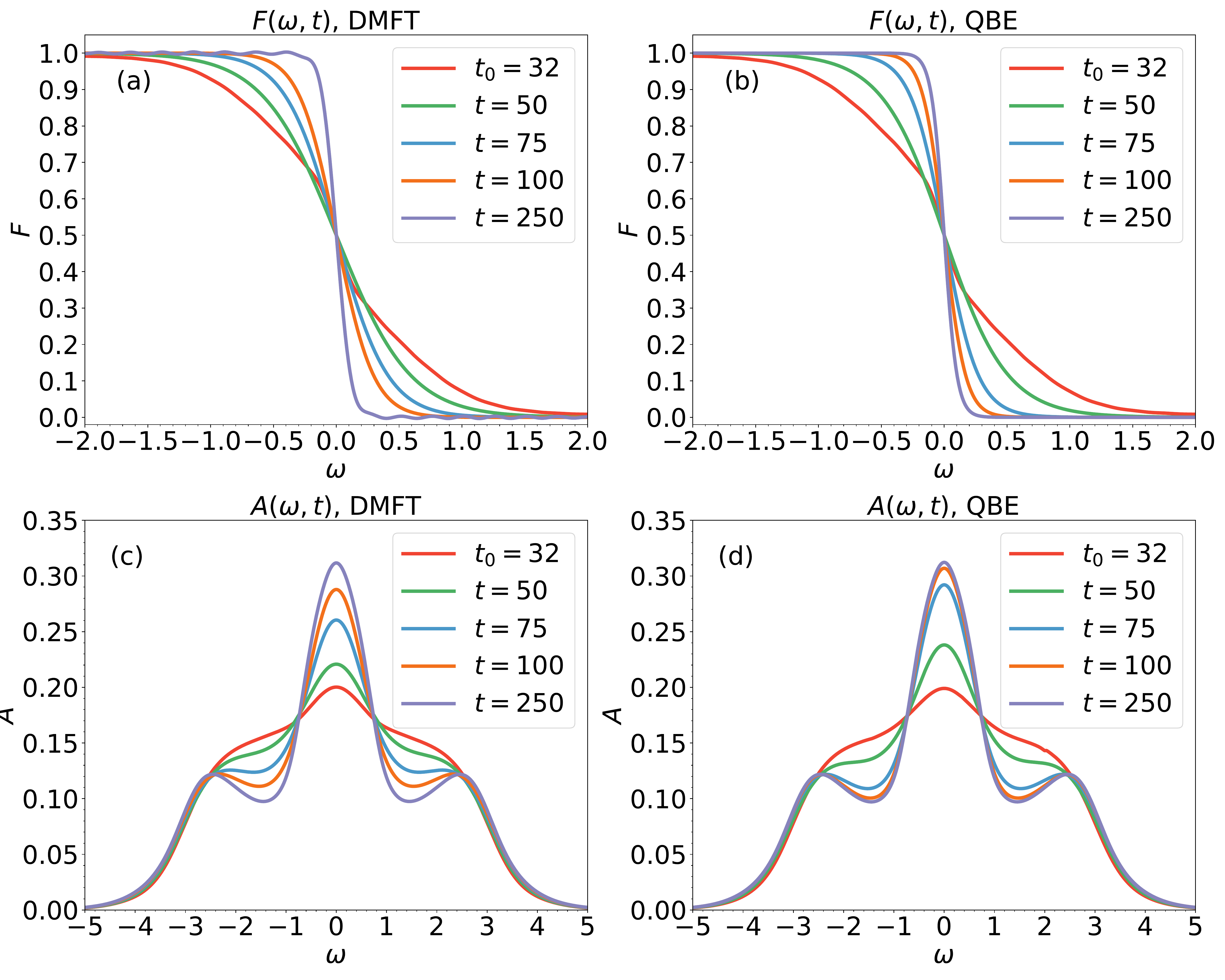}}
	\caption{
	Distribution function (upper panels) and spectral function (lower panels) obtained from the full DMFT solution (left panels) and the QBE (right panels) one at $U=3$. The QBE takes the DMFT distribution function $F(\omega,t_0)$ at time $t_0=32$ as initial state for the evolution at $t>t_0$ (initial inverse temperature $\beta=20$, electron phonon coupling $g^2=0.5$).}
	\label{fig03}
\end{figure}

Although the relaxation rate for the energy in the QBE seems to be overestimated for larger values of $U$, Fig. \ref{fig03} shows that the spectra and distribution functions obtained from the full DMFT and the QBE follow the same qualitative behavior, i.e., a relaxation of $F(\omega,t)$ to a Fermi function together with an evolution of the temperature in this  Fermi function towards the initial temperature. 
\section{Conclusion}
\label{Sec-IV}

In conclusion, we 
developed 
a kinetic equation which works without the need to assume the existence of quasiparticles with well-defined dispersion, $\epsilon_{\bm k}$ and, above all, evaluates  the scattering integral in a
non-perturbative
manner. In particular, a scattering integral which is consistent with DMFT is obtained by extracting self-energies from a quantum impurity model in an auxiliary non-equilibrium  steady state. Most importantly, this guaranties that the final state of the evolution is a proper description of the fully interacting state of the correlated electron system, which makes the present formalism unique with respect to conventional quantum kinetic approaches based on perturbative scattering integrals or certain assumptions on the spectral function, such as assuming a rigid density of states or the quasiparticle approximation. While for full non-equilibrium Green's function simulations  the numerical effort for the propagation over a time interval $t_\text{max}$ scales with $\mathcal{O}(t_\text{max}^3)$, and the required memory scales with $\mathcal{O}(t_\text{max}^2)$, in the QBE the numerical effort is linear with $t_\text{max}$ and the memory required is independent of $t_\text{max}$.

We have tested the framework on the relaxation of the electronic state in a correlated 
metal
after a population transfer that 
simulates
a photo-excitation. 
One 
assumption of the QBE, i.e., that the spectra at the correlated system can be obtained from an auxiliary steady state, is 
found to be satisfied
with remarkable accuracy.
Moreover,
the relaxation dynamics for both spectral functions and distribution functions within the full non-equilibrium DMFT simulation and the QBE are consistent. 
Quantitatively, the gradient approximation underlying the QBE leads to a slight overestimation of the relaxation rate. 
Whether this can be corrected by higher order expansions of the 
gradient approximation is left for future investigations.

The success of the QBE approach for the present setting motivates an application to different models. 
In particular this includes 
symmetry-broken states where interesting long-time phenomena have been observed,\cite{Picano2020} and the evolution of the Mott phase, where already a QBE with an ad-hoc scattering integral has shown relative success.\cite{Wais2018} Possible applications of the formalism include the evolution of the density of states in correlated systems, in particular multi-orbital systems where 
a
pronounced effect of the redistribution of weight has already been discussed using 
quasiparticle kinetic equations.\cite{He2015}  
In this context, the method can be combined with GW \cite{Wolf-PRL2014} or DMFT+GW\cite{Golez2019}, which have demonstrated again a pronounced dependence of the spectra on the distribution. 
Finally, another 
interesting perspective of the approach is that there are several promising numerical approaches to study  non-equilibrium steady states within DMFT. This includes variants of the strong-coupling expansion \cite{Scarlatella2019,Li2020b}, matrix product states,\cite{Schwarz2018} auxiliary master equations,\cite{Arrigoni2013} or Quantum Monte Carlo.\cite{Profumo2015,Bertrand2019} The QBE formalism would allow these non-perturbative techniques to access not only true steady states, but also non-equilibrium  states of correlated electrons on the picosecond timescale relevant for photo-induced phase transition and collective orders.

\acknowledgements
We acknowledge  Philipp Werner for useful discussions, and  Nagamalleswararao Dasari for discussions as well as his contribution to the implementation of the Ohmic bath.
This work was supported by the ERC Starting Grant No. 716648. The calculations 
have been done at the RRZE of the University Erlangen-Nuremberg.

\bibliography{biblio}
\end{document}